\documentstyle[12pt]{article}

\begin{document}
\begin{titlepage}
\setcounter{page}{1}
\title{Euclidean scalar and spinor \\ Green's functions
in Rindler space}
\author{B. LINET \thanks{E-mail: linet@ccr.jussieu.fr} \\
\mbox{\small Laboratoire de Gravitation et Cosmologie Relativistes} \\
\mbox{\small CNRS/URA 769, Universit\'{e} Pierre et Marie Curie} \\
\mbox{\small Tour 22/12, Bo\^{\i}te Courrier 142} \\
\mbox{\small 4, Place Jussieu, 75252 PARIS CEDEX 05, France}}
\maketitle
\begin{abstract}
In Rindler space, we consider the Feynman Green's functions
associated with either the Fulling-Rindler vacuum or the Minkowski vacuum.
In Euclidean field theory, they become
respectively the Euclidean Green's functions
$G_{\infty}$ and $G_{2\pi}$  whose we give different suitable forms.
In the case of the massive spin-$\frac{1}{2}$ field, we determine also the
Euclidean spinor Green's functions $S_{\infty}$ and  $S_{2\pi}$ in
different suitable forms. In both cases for massless fields in four
dimensions, we compute the vacuum expectation value of the
energy-momentum tensor relative to the Rindler observer.

\end{abstract}

\thispagestyle{empty}
\end{titlepage}

\section{Introduction}

We consider a Rindler observer in an $n$-dimensional Minkowski spacetime
$(n\geq 2)$ which has uniform acceleration, $g$.
In the coordinate system
$(\xi^{0},\xi ,x^{i})$, $i=1,..,n-2$, with $\xi >0$,
Rindler space is described by the metric
\begin{equation}
\label{1.1}
ds^{2}=-g^{2}\xi^{2}(d\xi^{0})^{2}+(d\xi)^{2}+(dx^{1})^{2}+
\cdots +(dx^{n-2})^{2}
\end{equation}
This coordinate system covers only a part of the Minkowski spacetime.
The line given by $\xi =1/g$ and $x^{i}=0$ corresponds to the world line of
the Rindler observer undergoing a uniform acceleration, $g$.

The free quantum field theory in Rindler space has been initially
investigated by Fulling \cite{fulling},
Davies \cite{davies}, Unruh \cite{unruh} and also Candelas and
Raine \cite{candelas} and Dowker \cite{dowker2}.
Due to the existence of the event horizon located at $\xi =0$, the quantum
field theory described in the Rindler metric is not equivalent to the
usual field theory in the Minkowski metric.
The Rindler observer moving in the Minkowskian vacuum sees a thermal
bath of temperature $g/2\pi$. We use units in which $c=\hbar =k_{B}=1$.

We recap briefly the definition of the Feynman Green's functions in terms of
Rindler metric (\ref{1.1}). Apart from the usual Minkowski vacuum
$\mid {\rm O}_{M}>$, there also exists the Fulling-Rindler vacuum
$\mid {\rm O}_{R}>$ which is associated with the timelike Killing vector
$\partial /\partial \xi^{0}$. Consequently, for a massive scalar field
of mass $m$, we have the two corresponding Feynman Green's functions
$\triangle_{R}$ and $\triangle_{M}$, defined respectively by
\begin{equation}
\label{1.2}
\triangle_{R}^{(n)}(x,x_{0};m)=<{\rm O}_{R}\mid {\rm T}
\varphi (x_{0})\varphi^{\dag}(x)\mid {\rm O}_{R}>
\end{equation}
\begin{equation}
\label{1.3}
\triangle_{M}^{(n)}(x,x_{0};m)=<{\rm O}_{M}\mid {\rm T}
\varphi (x_{0})\varphi^{\dag}(x)\mid {\rm O}_{M}>
\end{equation}
where T denotes the time-ordered product. We delete the factor $i$ and
futhermore we choose to renormalise the vacua to one.

By performing a Wick rotation involving the acceleration $g$ given by
\begin{equation}
\label{1.4}
\xi^{0}=-i\frac{\tau}{g}
\end{equation}
metric (\ref{1.1}) takes a Riemannian form
\begin{equation}
\label{1.5}
ds^{2}=\xi^{2}(d\tau )^{2}+(d\xi )^{2}+(dx^{1})^{2}+
\cdots +(dx^{n-2})^{2}
\end{equation}
in the coordinate system $(\tau ,\xi ,x^{i})$, $\xi >0$.
The properties of the Feynman Green's functions in a static spacetime
\cite{wald} imply that they can be deduced by an analytic continuation from
the Euclidean Green's functions. In the metric
(\ref{1.5}), we call respectively  $G_{\infty}$ and  $G_{2\pi}$
the corresponding Feynman Green's functions $\triangle_{R}$ and
$\triangle_{M}$.

These Green's functions have already been considered in previous
works but in the present paper we use the methods that we
introduced in the analysis of the Euclidean Green's functions in the
spacetime describing a straight cosmic string \cite{linet,guimaraes}.
We can then derive a convenient form of both $G_{\infty}$ and $G_{2\pi}$
in order to straightforwardly compute the vacuum expectation value of
the operators relative to the Rindler observer.

The plan of this work is as follows. In section 2, we determine the
Euclidean Green's functions $G_{2\pi}$ and $G_{\infty}$.
We give different forms of the expression for $G_{\infty}$ in section 3,
in particular a convenient expression for
$G_{2\pi}-G_{\infty}$ which is valid for point $x$ near $x_{0}$ is obtained.
We obtain the Euclidean spinor Green's functions $S_{2\pi}$ and
$S_{\infty}$ in section 4, in particular a convenient expression of
$S_{2\pi}-S_{\infty}$ valid for point $x$ near $x_{0}$.
We compute in section 5 the vacuum expectation value of the
energy-momentum tensor relative to the Rindler observer
for a massless scalar or spin-$\frac{1}{2}$ field in four dimensions.
In section 6, we introduce the Euclidean Green's function $G_{\beta}$ and
$S_{\beta}$ corresponding to an arbitrary temperature $1/\beta$.

\section{Scalar Green's functions}

The Feynman Green's function $\triangle_{R}$ is easily constructed from the
positive-frequency functions with respect to the coodinate $\tau$ of the
metric (\ref{1.1}); one finds that
\begin{eqnarray}
\label{2.1}
\nonumber & &\triangle_{R}^{(n)}(x,x_{0};m)=
\int d^{n-2}k\frac{1}{(2\pi )^{n-2}}
\exp [ik^{i}(x^{i}-x_{0}^{i})] \times \\
& &\int_{0}^{\infty}d\nu \frac{\sinh \pi \nu}{\pi^{2}}
\exp [-i\nu (g\mid \xi^{0}-\xi_{0}^{0}\mid -i\epsilon )]
K_{i\nu}(\kappa_{k}\xi )K_{i\nu}(\kappa_{k}\xi_{0})
\end{eqnarray}
where $\kappa_{k}=\sqrt{k^{i}k^{i}+m^{2}}$, in the limit
$\epsilon \rightarrow 0$.

The Feynman Green's function, $\triangle_{M}$ is calculated from the
development of the creation and annihilation operators associated with the
Fulling-Rindler vacuum on the Minkowskian creation and annhilation
operators ; Spindel \cite{spindel} gives
the following expression
\begin{eqnarray}
\label{2.2}
\nonumber & &\triangle_{M}^{(n)}(x,x_{0};m)=
\int d^{n-2}k\frac{1}{(2\pi )^{n-2}}
\exp [ik^{i}(x^{i}-x_{0}^{i})] \times  \\
& &\int_{0}^{\infty}d\nu \frac{1}{\pi^{2}}\cosh [\nu (\pi -ig\mid \xi^{0}-
\xi_{0}^{0}\mid -\epsilon )]K_{i\nu}(\kappa_{k}\xi )
K_{i\nu}(\kappa_{k}\xi_{0})
\end{eqnarray}
in the limit $\epsilon \rightarrow 0$.

The Wick rotation (\ref{1.4}) applied to expressions (\ref{2.1})
and (\ref{2.2}) yields the following expression for the Green's function,
$G_{\infty}$
\begin{eqnarray}
\label{2.3}
\nonumber & &G_{\infty}^{(n)}(x,x_{0};m)=
\int d^{n-2}k\frac{1}{(2\pi )^{n-2}}
\exp ik^{i}(x^{i}-x_{0}^{i}) \times \\
& &\int_{0}^{\infty}d\nu \frac{\sinh \pi \nu}{\pi^{2}}
\exp [-\nu \mid \tau -\tau_{0}\mid ]K_{i\nu}(\kappa_{k}\xi)
K_{i\nu}(\kappa_{k}\xi_{0})
\end{eqnarray}
and also for the Green's function, $G_{2\pi}$
\begin{eqnarray}
\label{2.4}
\nonumber & &G_{2\pi}^{(n)}(x,x_{0};m)=\int d^{n-2}k\frac{1}{(2\pi )^{n-2}}
\exp [ik^{i}(x^{i}-x_{0}^{i})] \times \\
& &\int_{0}^{\infty}d\nu \frac{1}{\pi^{2}}
\cosh [\nu (\pi-\mid \tau -\tau_{0}\mid )]K_{i\nu}(\kappa_{k}\xi )
K_{i\nu}(\kappa_{k}\xi_{0})
\end{eqnarray}
In fact, the multiple integral (\ref{2.4}) can be explicitly
calculated; one has merely
\begin{equation}
\label{2.5}
G_{2\pi}^{(n)}(x,x_{0};m)=\frac{m^{n/2-1}}{(2\pi )^{n/2}r_{n}^{n/2-1}}
K_{n/2-1}(mr_{n})
\end{equation}
where $r_{n}=\sqrt{\xi^{2}+\xi_{0}^{2}-2\xi \xi_{0}\cos (\tau -\tau_{0})
+(x^{1}-x_{0}^{1})^{2}+\cdots (x^{n-2}-x_{0}^{n-2})^{2}}$.
Evidently, $G_{2\pi}$ is periodic in the coordinate $\tau$ with
period $2\pi$.

Within Euclidean field theory, one must impose that the manifold defined
by the metric (\ref{1.5}) is regular. Hence, we require the coodinate $\tau$
to be a periodic coordinate ranging from 0 to $2\pi$. Thus, it coincides
with the Euclidean space.
The Green's function $G_{2\pi}$ is now the ordinary
Green's function in an Euclidean space.
{}From expressions (\ref{2.3}) and (\ref{2.4}), one can prove that
\begin{equation}
\label{2.6}
G_{2\pi}(\tau -\tau_{0})=\sum_{n=-\infty}^{\infty}
G_{\infty}(\tau -\tau_{0}+2\pi n)
\end{equation}
in which the dependence of the space coordinates is not indicated.
In the quantum field theory at finite temperature, we can thus interpret
$G_{2\pi}$ as the thermal Euclidean Green's function
\cite{israel,christensen,gibbons} with respect to
$G_{\infty}$ which is the zero-temperature Green's function. Considered
in the spacetime (\ref{1.1}), the periodicity of $G_{2\pi}$ in the
coordinate $\xi^{0}$ is an imaginary period $i2\pi /g$. This is the origin
of the thermal character of the Minkowski vacuum perceived by the Rindler
observer at temperature $T_{0}=g/2\pi$.

The vacuum expectation value of the operators relative to the Rindler
observer is obtained from the Green's function $G_{2\pi}$, but the
renormalisation is performed by removing the Green's function $G_{\infty}$.
As in the Casimir effect, $G_{\infty}$ plays the role
of the local Green's function and $G_{2\pi}$ the global Green's function
\cite{brown}. In a symbolic manner, we write
\begin{equation}
\label{2.7}
<O(x)>={\cal O}(G_{2\pi}-G_{\infty})\mid_{x=x_{0}}
\end{equation}
where ${\cal O}$ is a differentiel operator in $x$ and $x_{0}$.

\section{Determination of $G_{\infty}$}

We now search expressions for the Green's function $G_{\infty}$ which are
more suitable than that given by equation (\ref{2.3}). We start from $n=2$
because we have the
following recurrence relation between the Green's functions
\begin{eqnarray}
\label{3.1}
\nonumber & &G_{\infty}^{(n)}(\tau ,\xi ,x^{1},\cdots ,x^{n-2};m)=
\frac{1}{2\pi} \times  \\
& &\int_{-\infty}^{\infty}dkG_{\infty}^{(n-1)}(\tau ,\xi ,x^{1},\cdots
,x^{n-3}; \sqrt{m^{2}+k^{2}})\cos [k(x^{n-2}-x_{0}^{n-2})]
\end{eqnarray}
for $n\geq 3$.

\subsection{Integral expression of $G_{\infty}$}

For this purpose, we make use of the formula
\[
\sinh (\pi \nu )K_{i\nu}(m\xi )K_{i\nu}(m\xi_{0})=
\frac{\pi}{2}\int_{\xi_{2}}^{\infty}duJ_{0}[m(2\xi \xi_{0})^{1/2}
(\cosh u-\cosh \xi_{2})^{1/2}]\sin (\nu u)
\]
where $\xi_{2}$ is defined by
\[
\cosh \xi_{2}=\frac{\xi^{2}+\xi_{0}^{2}}{2\xi \xi_{0}} \quad (\xi_{2}\geq 0)
\]
By inserting this into (\ref{2.3}) for $n=2$, we have
\begin{eqnarray}
\nonumber & &G_{\infty}^{(2)}(x,x_{0};m)=\frac{1}{2\pi}\int_{0}^{\infty}d\nu
\exp [-\nu \mid \tau -\tau_{0}\mid ] \times \\
\nonumber & &\int_{\xi_{2}}^{\infty}duJ_{0}[m(2\xi \xi_{0})^{1/2}
(\cosh u-\cosh \xi_{2})^{1/2}]\sin (\nu u)
\end{eqnarray}
Taking into account the formula
\[
\int_{0}^{\infty}dx\exp (-px)\sin (qx)=\frac{q}{p^{2}+q^{2}} \quad  (p>0)
\]
we find an integral expression of $G_{\infty}^{(2)}$ of the form
\begin{eqnarray}
\label{3.2}
\nonumber & &G_{\infty}^{(2)}(x,x_{0};m)= \frac{1}{2\pi} \times \\
& &\int_{\xi_{2}}^{\infty}duJ_{0}[m(2\xi \xi_{0})^{1/2}
(\cosh u-\cosh \xi_{2})^{1/2}]\frac{u}{u^{2}+(\tau -\tau_{0})^{2}}
\end{eqnarray}

By virtue of the recurrence relation (\ref{3.1}), we obtain in three
dimensions
\begin{eqnarray}
\label{3.3}
\nonumber & &G_{\infty}^{(3)}(x,x_{0};m)=\frac{1}{2\pi^{2}(2\xi
\xi_{0})^{1/2}}    \times \\
& &\int_{\xi_{3}}^{\infty}du
\frac{\cos [m(2\xi \xi_{0})^{1/2}(\cosh u-\cosh \xi_{3})^{1/2}]}
{(\cosh u-\cosh \xi_{3})^{1/2}}\frac{u}{u^{2}+(\tau -\tau_{0})^{2}}
\end{eqnarray}
where $\xi_{3}$ is defined by
\[
\cosh \xi_{3}=\frac{\xi^{2}+\xi_{0}^{2}+(x^{1}-x_{0}^{1})^{2}}
{2\xi \xi_{0}}    \quad (\xi_{3} \geq 0)
\]
Upon using again (\ref{3.1}), we obtain in four dimensions
\begin{eqnarray}
\label{3.4}
\nonumber & &G_{\infty}^{(4)}(x,x_{0};m)=-\frac{1}{4\pi^{2}\xi \xi_{0}}
\int_{\xi_{4}}^{\infty}du
J_{0}[m(2\xi \xi_{0})^{1/2}(\cosh u-\cosh \xi_{4})^{1/2}] \times \\
& &\frac{d}{du}[\frac{u}{\sinh u(u^{2}+(\tau -\tau_{0})^{2})}]
\end{eqnarray}
where $\xi_{4}$ is defined by
\[
\cosh \xi_{4}=\frac{\xi^{2}+\xi_{0}^{2}+(x^{1}-x_{0}^{1})^{2}
+(x^{2}-x_{0}^{2})^{2}}{2\xi \xi_{0}}   \quad (\xi_{4}\geq 0)
\]
and so on as in \cite{guimaraes}. We emphasize that (\ref{3.4}) can be
explicitly integrated in the case $m=0$; we find
\begin{equation}
\label{3.5}
D_{\infty}^{(4)}(x,x_{0})=\frac{\xi_{4}}{4\pi^{2}\xi \xi_{0}\sinh \xi_{4}
(\xi_{4}^{2}+(\tau -\tau_{0})^{2})}
\end{equation}
which agrees with the result of Troost and Van Dam \cite{troost}.

\subsection{Local expression of $G_{\infty}$}

As explained in section 2, the calculation of the vacuum
expectation value is performed in the coincidence limit $x=x_{0}$.
According to the formula (\ref{2.7}), we must know $G_{2\pi}-G_{\infty}$
for point $x$ near point $x_{0}$.
We anticipate and we say that the required subset of the Rindler manifold
is defined by the condition
\begin{equation}
\label{3.6}
\mid \tau -\tau_{0}\mid <\pi
\end{equation}
satisfied of course when $x$ tends to $x_{0}$.

For this purpose, we make use of the formula
\[
K_{i\nu}(m\xi )K_{i\nu}(m\xi_{0})=\int_{0}^{\infty}duK_{0}[mR_{2}(u)]
\cos (\nu u)
\]
where $R_{2}(u)=\sqrt{\xi^{2}+\xi_{0}^{2}+2\xi \xi_{0}\cosh u}$.
Insertion of this into (\ref{2.3}) for $n=2$ yields
\begin{eqnarray}
\nonumber & &G_{\infty}^{(2)}(x,x_{0};m)=\frac{1}{\pi^{2}}  \times \\
\nonumber & &\int_{0}^{\infty}d\nu \sinh (\pi \nu )
\exp [-\nu \mid \tau -\tau_{0}\mid ]
\int_{0}^{\infty}duK_{0}[mR_{2}(u)]\cos (\nu u)
\end{eqnarray}
Due to the identity
\[
\sinh (\pi \nu )\exp (-\nu \psi)=\cosh [\nu (\psi -\pi )]-\cosh (\nu
\psi )\exp (-\pi \nu )
\]
we can rewrite $G_{\infty}^{(2)}$ as
\begin{eqnarray}
\label{3.7}
\nonumber & &G_{\infty}^{(2)}(x,x_{0};m)=G_{2\pi}^{(2)}(x,x_{0};m)
-\frac{1}{\pi^{2}} \times    \\
& &\int_{0}^{\infty}duK_{0}[mR_{2}(u)]
\int_{0}^{\infty}d\nu \cosh [\nu \mid \tau -\tau_{0}\mid ]
\exp (-\pi \nu )\cos (\nu u )
\end{eqnarray}
in which the integral converges under assumption (\ref{3.6}).
The $\nu$-integration in (\ref{3.7}) can be performed with the aid of the
formula
\[
\int_{0}^{\infty}d\nu \cosh (\nu \psi )\exp (-\pi \nu )\cos (\nu u)=
\frac{1}{2}[\frac{-\psi +\pi}{(\psi -\pi )^{2}+u^{2}}+
\frac{\psi +\pi}{(\psi +\pi )^{2}+u^{2}}]
\]
Finally, we obtain the desired local expression of $G_{\infty}^{(2)}$
\begin{eqnarray}
\label{3.8}
\nonumber & &G_{\infty}^{(2)}(x,x_{0};m)=G_{2\pi}^{(2)}(x,x_{0};m)+
\frac{1}{4\pi^{2}}  \times \\
& &\int_{0}^{\infty}duK_{0}[mR_{2}(u)]F_{\infty}(u,\tau -\tau_{0})
\end{eqnarray}
where the function $F_{\infty}(u,\psi )$, which is well defined everywhere,
is given by
\[
F_{\infty}(u,\psi )=2[-\frac{\pi +\psi}{(\pi +\psi )^{2}+u^{2}}
+\frac{\psi -\pi}{(\psi -\pi )^{2}+u^{2}}]
\]

By virtue of the recurrence relation (\ref{3.1}), we obtain
in three dimensions
\begin{eqnarray}
\label{3.9}
\nonumber & &G_{\infty}^{(3)}(x,x_{0};m)=G_{2\pi}^{(3)}(x,x_{0};m)+
\frac{1}{8\pi^{2}} \times \\
& &\int_{0}^{\infty}du\frac{\exp [-mR_{3}(u)]}{R_{3}(u)}
F_{\infty}(u,\tau -\tau_{0})
\end{eqnarray}
where $R_{3}(u)=\sqrt{R_{2}^{2}(u)+(x^{1}-x_{0}^{1})^{2}}$.
In four dimensions, we obtain
\begin{eqnarray}
\label{3.10}
\nonumber & &G_{\infty}^{(4)}(x,x_{0};m)=G_{2\pi}^{(4)}(x,x_{0};m)+
\frac{m}{8\pi^{3}} \times \\
& &\int_{0}^{\infty}du\frac{K_{1}[mR_{4}(u)]}{R_{4}(u)}
F_{\infty}(u,\tau -\tau_{0})
\end{eqnarray}
where $R_{4}(u)=\sqrt{R_{3}^{2}(u)+(x^{2}-x_{0}^{2})^{2}}$ and so on
as in \cite{guimaraes}, again in the subset defined by
condition (\ref{3.6}). In the case $m=0$, expression (\ref{3.10})
reduces to
\begin{eqnarray}
\label{3.11}
\nonumber & &D_{\infty}^{(4)}(x,x_{0})= D_{2\pi}^{(4)}(x,x_{0})+
\frac{1}{8\pi^{3}}  \times \\
& &\int_{0}^{\infty}du\frac{1}{[R_{4}(u)]^{2}}F_{\infty}(u,\tau -\tau_{0})
\end{eqnarray}

\section{Spinor Green's functions}

In the quantum field theory for a spin-$\frac{1}{2}$ field in the
Rindler spacetime, the Rindler observer also sees a thermal bath of
temperature $g/2\pi$
\cite{candelas2,soffel,hacyan}. It is possible to use the Euclidean approach
and in particular the theory at finite temperature in which the Rindler
manifold is regular. In this framework, we must determine the Euclidean
spinor Green's functions $\overline{S}_{\infty}$ et $\overline{S}_{2\pi}$,
expressed in a vierbein well defined as that associated with the
Cartesian coordinates, which satisfy the two conditions
\begin{equation}
\label{7.1}
\overline{S}_{2\pi}(\tau -\tau_{0}+2\pi )=-\overline{S}_{2\pi}
(\tau -\tau_{0})
\end{equation}
and
\begin{equation}
\label{7.2}
\overline{S}_{2\pi}(\tau -\tau_{0})=\sum_{n=-\infty}^{\infty}(-1)^{n}
\overline{S}_{\infty}(\tau -\tau_{0}+2\pi n)
\end{equation}
In this interpretation, $\overline{S}_{2\pi}$ represents the
finite temperature spinor Green's function at the temperature $T_{0}=g/2\pi$
and $\overline{S}_{\infty}$ the zero-temperature Green's function.
We point out that $\overline{S}_{2\pi}$ is not the ordinary spinor
Green's function $S_{E}$ in Euclidean space but it corresponds to the
twisted spinor Green's function $S_{E}^{(T)}$.

However in metric (\ref{1.5}), we defines  another vierbein
$e_{\underline{a}}^{\mu}$ which is given
by the components
\begin{equation}
\label{6.1}
e_{\underline{1}}^{\mu}=(\frac{1}{\xi},0,...,0) \quad {\rm et}
\quad e_{\underline{a}}^{\mu}=\delta_{\underline{a}}^{\mu} \quad
\underline{a}=2,...,n
\end{equation}
The spinor connection $\Gamma_{\mu}$ then has the components
\begin{equation}
\label{6.2}
\Gamma_{1}=-\frac{1}{4}(\gamma^{\underline{2}}\gamma^{\underline{1}}
-\gamma^{\underline{1}}\gamma^{\underline{2}}) \quad  {\rm et} \quad
\Gamma_{i}=0 \quad i=2,...,n
\end{equation}
Taking into account the transformation laws of the spinor components,
the spinor Green's functions can be expressed in terms of vierbein
(\ref{6.1})
in the notations $S_{\infty}$ and $S_{2\pi}$. With this choice of
vierbein, condition (\ref{7.1}) takes the following form on the
spinor Green's function $S_{2\pi}$
\begin{equation}
\label{6.3}
S_{2\pi}(\tau -\tau_{0}+2\pi )=S_{2\pi}(\tau -\tau_{0})
\end{equation}
and condition (\ref{7.2}) becomes
\begin{equation}
\label{6.3a}
S_{2\pi}(\tau -\tau_{0})=\sum_{n=-\infty}^{\infty}S_{\infty}
(\tau -\tau_{0}+2\pi n)
\end{equation}

In $n$ dimensions ($n\geq 2$),  we have determined in a previous work
\cite{linet2} the spinor Green's function
satisfying (\ref{6.3}) in the notation $S^{(n)(T)}$ for the value $B=1$.
We recall the result
\begin{equation}
\label{6.4}
S_{2\pi}(x,x_{0};m)=(e_{\underline{a}}^{\mu}\gamma^{\underline{a}}
\partial_{\mu}+\frac{\gamma^{\underline{2}}}{2\xi}-mI)[
I\Re H_{2\pi}+\gamma^{\underline{2}}\gamma^{\underline{1}}\Im H_{2\pi}]
\end{equation}
where
\[
H_{2\pi}(x,x_{0};m)=\exp i\frac{\tau -\tau_{0}}{2}G_{2\pi}^{(T)}(x,x_{0};m)
\]
and where $G_{2\pi}^{(T)}$ is the scalar Green's function satisfying
\begin{equation}
\label{6.5}
G_{2\pi}^{(T)}(\tau -\tau_{0}+2\pi )=-G_{2\pi}^{(T)}(\tau -\tau_{0})
\end{equation}
In consequence, we obviously have
\begin{equation}
\label{6.6}
G_{2\pi}^{(T)}(\tau -\tau_{0})=\sum_{n=-\infty}^{\infty}
(-1)^{n}G_{\infty}(\tau -\tau_{0}+2\pi n)
\end{equation}
where $G_{\infty}$ is the scalar Green's function given by expression
(\ref{2.3}) in section 3.2. Consequently, we can define the spinor Green's
function $S_{\infty}$ by
\begin{equation}
\label{6.7}
S_{\infty}(x,x_{0};m)=(e_{\underline{a}}^{\mu}\gamma^{\underline{a}}
\partial_{\mu}+\frac{\gamma^{\underline{2}}}{2\xi}-mI)[
I\Re H_{\infty}+\gamma^{\underline{2}}\gamma^{\underline{1}}\Im H_{\infty}]
\end{equation}
with
\[
H_{\infty}(x,x_{0};m)=\exp i\frac{\tau -\tau_{0}}{2}G_{\infty}(x,x_{0};m)
\]
{}From relation (\ref{6.6}), we deduce the property (\ref{6.3a})  which
corresponds to the fundamental property (\ref{7.2}).

As in the case of the scalar field, we calculate the vacuum expectation
value relative to the Rindler observer by taking $S_{2\pi}-S_{\infty}$
in the coincidence limit. In a symbolic manner, we write
\begin{equation}
\label{6.8}
<O(x)>={\cal O}^{(1/2)}(S_{2\pi}-S_{\infty})\mid_{x=x_{0}}
\end{equation}
where ${\cal O}^{(1/2)}$ is a differential operator in $x$ and $x_{0}$.

We now turn our attention to the derivation of an expression for
$S_{2\pi}-S_{\infty}$ which is
valid for a point $x$ in the neighbourhood of $x_{0}$
that we can look for in the form
\[
S_{2\pi}-S_{\infty}=(S_{2\pi}-S_{E})-(S_{\infty}-S_{E})
\]
Taking into account the formulae
(\ref{6.4}) and (\ref{6.7}), we can thus obtain the required expression
by finding  $G_{2\pi}^{(T)}-G_{\infty}$ in the form
\[
G_{2\pi}^{(T)}-G_{\infty}=(G_{2\pi}^{(T)}-G_{2\pi})-(G_{\infty}-G_{2\pi})
\]
In a previous work \cite{guimaraes}, we have obtained in an arbitrary
number of dimensions such an expression for $G_{2\pi}^{(T)}-G_{2\pi}$,
where $G_{2\pi}^{(T)}$ is given under the notation $G_{1/2}^{(n)}$
for the value $B=1$. On the other hand,
the expression for $G_{\infty}-G_{2\pi}$ has been given in section 3.2.
So, we can determined the desired expression of $G_{2\pi}^{(T)}-G_{\infty}$.

We confine ourselves to considering $S_{2\pi}^{(4)}$  and $S_{\infty}^{(4)}$
in four dimensions. We recall the expression for
$G_{2\pi}^{(4)(T)}$ subject to the condition (\ref{3.6})
\begin{eqnarray}
\label{6.11}
\nonumber & &G_{2\pi}^{(4)(T)}(x,x_{0};m)=G_{2\pi}^{(4)}(x,x_{0};m)
+\frac{m}{8\pi^{3}} \times \\
& &\int_{0}^{\infty}du
\frac{K_{1}[mR_{4}(u)]}{R_{4}(u)}F^{(1/2)}_{1}(u,\tau -\tau_{0})
\end{eqnarray}
where the function $F^{(1/2)}_{1}(u,\psi )$ is given by
\[
F^{(1/2)}_{1}(u,\psi )=-4\cosh \frac{u}{2}\cos \frac{\psi}{2}
\frac{1}{\cosh u+\cos \psi}
\]
Taking into account expression (\ref{3.10}), we obtain
\begin{eqnarray}
\label{6.15}
\nonumber & &G_{2\pi}^{(4)(T)}(x,x_{0};m)-G_{\infty}^{(4)}(x,x_{0};m)=
\frac{m}{8\pi^{3}}\int_{0}^{\infty}du
\frac{K_{1}[mR_{4}(u)]}{R_{4}(u)} \times \\
& &[F^{(1/2)}_{1}(u,\tau -\tau_{0})-F_{\infty}(u,\tau -\tau_{0})]
\end{eqnarray}
which is valid under condition (\ref{3.6}).
In the case $m=0$, expression (\ref{6.15}) reduces to
\begin{eqnarray}
\label{6.16}
\nonumber & &D_{2\pi}^{(4)(T)}(x,x_{0})-D_{\infty}^{(4)}(x,x_{0})=
\frac{1}{8\pi^{3}}\int_{0}^{\infty}du\frac{1}{[R_{4}(u)]^{2}} \times \\
& &[F^{(1/2)}_{1}(u,\tau -\tau_{0})-F_{\infty}(u,\tau -\tau_{0})]
\end{eqnarray}
which is, of course, valid under condition (\ref{3.6}).

\section{Vacuum energy-momentum tensor $(n=4)$}

We intend to compute the vacuum expectation value of the
energy-momentum tensor relative to the Rindler observer in the
coordinate system $(\tau ,\xi ,x^{i})$ with the help of formulae
(\ref{2.7}) and (\ref{6.8}). Since Rindler space is static
we can then deduce from this the vacuum energy-momentum tensor
in the coordinate system $(\xi^{0},\xi ,x^{i})$
by using the Wick rotation (\ref{1.4}).

We remark that we can also express these results in terms of a local
temperature $T$ given by
\begin{equation}
\label{9.1}
T=\frac{T_{0}}{g\xi} \quad {\rm or} \quad T=\frac{1}{2\pi\xi}
\end{equation}
for the value $T_{0}=g/2\pi$. It is convenient for all Rindler observers
$\xi =const.$ and $x^{i}=0$.

\subsection{Case of the scalar field}

At first, we calculate the mean-square field $<\phi^{2}>$ using the formula
\begin{equation}
\label{9.2}
<\phi^{2}(x)>=(G_{2\pi}^{(4)}(x,x_{0};m)-G_{\infty}^{(4)}(x,x_{0};m))
\mid_{x=x_{0}}
\end{equation}
{}From (\ref{3.10}), we obtain
\begin{equation}
\label{9.3}
<\phi^{2}(x)>=\frac{m}{4\pi^{2}}\int_{0}^{\infty}du
\frac{K_{1}[2m\xi \cosh (u/2)]}{\cosh (u/2)(\pi^{2}+u^{2})}
\end{equation}
For the case $m=0$, expression (\ref{9.3}) reduces to
\begin{equation}
\label{9.4}
<\phi^{2}(x)>=\frac{1}{8\pi^{2}\xi^{2}}\int_{0}^{\infty}du
\frac{1}{\cosh (u/2)(1+u^{2})}
\end{equation}
Upon applying the formula
\[
\int_{0}^{\infty}\frac{1}{(1+\cosh (\pi x))(1+x^{2})}=\frac{1}{12}\pi
\]
we have thereby
\begin{equation}
\label{9.5}
<\phi^{2}(x)>=\frac{1}{48\pi^{2}\xi^{2}}
\end{equation}

The vacuum expectation value of the energy-momentum tensor is given by
\begin{equation}
\label{9.10}
<T_{\mu}^{\nu}(x)>={\cal T}_{\mu}^{\nu}(G_{2\pi}(x,x_{0};m)-
G_{\infty}(x,x_{0};m))\mid_{x=x_{0}}
\end{equation}
where ${\cal T}_{\mu}^{\nu}$ is the following differential operator in $x$
and $x_{0}$
\begin{equation}
\label{9.11}
{\cal T}_{\mu}^{\nu}=(1-2\Xi )\nabla_{\mu}\nabla^{\nu_{0}}-2\Xi
\nabla_{\mu}\nabla^{\nu}+(2\Xi-\frac{1}{2})\delta_{\mu}^{\nu}
(m^{2}+\nabla_{\alpha}\nabla^{\alpha_{0}})
\end{equation}
with $\Xi$ the parameter of the scalar field theory under consideration.
The application of formula (\ref{9.10}) leads to well defined integrals.
As an example for a conformally invariant scalar field,
{\em i.e.} $m=0$ and $\Xi =1/6$, it is possible to integrate explicitly them.
Without given details, we find
\begin{equation}
\label{9.12}
<T_{\mu}^{\nu}(x)>=\frac{1}{1440\pi^{2}\xi^{4}}{\rm diag.}(-3,1,1,1)
\end{equation}
By the inverse of the Wick rotation (\ref{1.4}), we deduce
\begin{equation}
\label{9.13}
<T_{\xi^{0}}^{\xi^{0}}(x)>=-\frac{1}{480\pi^{2}\xi^{4}}
\end{equation}
which gives a positive vacuum energy density
according to the signature of metric (\ref{1.1}).

Results (\ref{9.5}) and (\ref{9.13}) can be also expressed in the form
\begin{equation}
\label{9.14}
<\phi^{2}(x)>=\frac{1}{12}T^{2} \quad {\rm et} \quad
<T_{\xi^{0}}^{\xi^{0}}(x)>=-\frac{\pi^{2}}{30}T^{4}
\end{equation}
where the temperature $T$ is given by (\ref{9.1}).

\subsection{Case of the spinor field}

The vacuum expectation value of the energy-momentum tensor is given by
\begin{equation}
\label{9.15}
<T_{\mu}^{\nu}(x)>={\cal T}^{(1/2)\nu}_{\quad \mu}(S_{2\pi}^{(4)}(x,x_{0};m)
-S_{\infty}^{(4)}(x,x_{0};m))\mid_{x=x_{0}}
\end{equation}
where ${\cal T}^{(1/2)\nu}_{\quad \mu}$ is the following differential
operator in $x$ and $x_{0}$
\begin{equation}
\label{9.16}
{\cal T}^{(1/2)}_{\mu \nu}=\frac{1}{4}tr[\gamma^{\underline{a}}(
e_{\underline{a}\mu}(\partial_{\nu}-\partial_{\nu_{0}})+
e_{\underline{a}\nu}(\partial_{\mu}-\partial_{\mu_{0}}))]
\end{equation}
The application of formula (\ref{9.15}) leads to well defined integrals.

We confine ourselves to the case $m=0$ in four dimensions.
We examine firstly the component $<T_{x^{1}}^{x^{1}}>$. From (\ref{9.15}),
we find
\begin{equation}
\label{9.17}
<T_{x^{1}}^{x^{1}}(x)>=4\partial_{x^{1}x^{1}}(H_{2\pi}^{(4)}(x,x_{0})-
H_{\infty}^{(4)}(x,x_{0}))\mid_{x=x_{0}}
\end{equation}
By substituting (\ref{6.16}) into (\ref{9.17}), we obtain
\[
<T_{x^{1}}^{x^{1}}(x)>=-\frac{1}{4\pi^{3}\xi^{4}}\int_{0}^{\infty}du
\frac{1}{(1+\cosh u)^{2}}[F^{(1/2)}_{1}(u,0)-F_{\infty}(u,0)]
\]
which we can write as
\[
<T_{x^{1}}^{x^{1}}(x)>=\frac{1}{\pi^{3}\xi^{4}}\int_{0}^{\infty}du
\frac{1}{(1+\cosh u)^{2}}[\frac{\cosh u/2}{1+\cosh u}-
\frac{\pi}{\pi^{2}+u^{2}}]
\]
By using the following integrals
\[
\int_{0}^{\infty}dx\frac{1}{(1+\cosh (\pi x))^{2}(1+x^{2})}=
\frac{11}{360}\pi
\]
\[
\int_{0}^{\infty}dx\frac{1}{\cosh^{5}x}=\frac{3}{16}\pi
\]
we obtain finally
\begin{equation}
\label{9.18}
<T_{x^{1}}^{x^{1}}(x)>=\frac{1}{\pi^{2}\xi^{4}}[\frac{3}{64}-\frac{11}{360}]
\end{equation}
Consequently, the vacuum energy-momentum relative to the Rindler
observer has the expression
\begin{equation}
\label{9.19}
<T_{\mu}^{\nu}(x)>=\frac{47}{2880\pi^{2}\xi^{4}}{\rm diag.}(-3,1,1,1)
\end{equation}
since it must be traceless and conserved.
Our result (\ref{9.19}) does not agree with the one due to Candelas
and Deutsch \cite{candelas2} and Dowker \cite{dowker4}.

\section{Fields at an arbitrary temperature}

In the metric (\ref{1.5}), the Euclidean scalar Green's function $G_{\beta}$
at finite temperature $T_{0}$ is periodic in the coordinate $\tau$
with period $\beta$, where $\beta =1/T_{0}$. When $\beta =2\pi$,
$G_{\beta}$ obviously coincides with $G_{2\pi}$. Moreover, $G_{\infty}$
is obtained as the limiting case of $G_{\beta}$ when $T_{0}$
tends to zero.

A transcript of our results \cite{guimaraes} on the scalar Green's function
for a massive scalar field in the spacetime describing a straight cosmic
string enable us to determine $G_{\beta}$. We must set
$\tau=B\varphi$ and $\beta =2\pi B$ in our formulae. We remark that,
in the massless case in four dimensions,
$D_{\beta}^{(4)}$  has been already given by Dowker \cite{dowker3}.

We have in particular found a local form of $G_{\beta}$  which is
convenient when one computes the vacuum
expectation value of the operators. We set
\begin{equation}
\label{10.1}
<O(x)>_{\beta}={\cal O}(G_{\beta}(x,x_{0};m)-
G_{\infty}(x,x_{0};m))\mid_{x=x_{0}}
\end{equation}
In the conformally invariant case, we find
\begin{equation}
\label{10.1a}
<T_{\mu}^{\nu}(x)>_{\beta}=\frac{\pi^{2}}{90\xi^{4}\beta^{4}}
{\rm diag.}(-3,1,1,1)
\end{equation}
When $\beta =2\pi$, we find again the energy-momentum tensor (\ref{9.12}).

The Euclidean spinor Green's function $\overline{S}_{\beta}$
is antiperiodic in $\tau$ with period $\beta$.
Likewise, a transcript of our results \cite{linet2} on the twisted
spinor Green's
function for a massive spin-$\frac{1}{2}$ field in the spacetime describing
a straight cosmic string enable us to determine $\overline{S}_{\beta}$,
or more precisely $S_{\beta}$ in the choosen vierbein.
We must set $\tau =B\varphi$ and $\beta =2\pi B$ in the expression for
the twisted spinor Green's function. We also set
\begin{equation}
\label{10.2}
<O(x)>_{\beta}={\cal O}^{(1/2)}(S_{\beta}(x,x_{0};m)-
S_{\infty}(x,x_{0};m))\mid_{x=x_{0}}
\end{equation}
We apply formula (\ref{10.2}) for the energy-momentum tensor in the
massless case by setting
\[
S_{\beta}-S_{\infty}=(S_{\beta}-S_{E})-(S_{\infty}-S_{E})
\]
Making use of previous calculations for $\beta >\pi$ \cite{linet2,dowker1},
we find
\begin{equation}
\label{10.4}
<T_{\mu}^{\nu}(x)>_{\beta}=\frac{1}{\xi^{4}}[-2w_{4}(\gamma )-
\frac{11}{360\pi^{2}}]{\rm diag.}(-3,1,1,1)
\end{equation}
where $w_{4}(\gamma )$ is the following expression
\begin{eqnarray}
\label{10.4a}
\nonumber & &w_{4}(\gamma )=-\frac{1}{720\pi^{2}}\{ 11 \\
& &-\frac{60\pi^{2}}{\beta^{2}}[
4(\gamma -\frac{1}{2})^{2}-\frac{1}{3}]+\frac{30\pi^{4}}{\beta^{4}}[
16(\gamma -\frac{1}{2})^{4}-8(\gamma -\frac{1}{2})^{2}+\frac{7}{15}]\}
\end{eqnarray}
where the parameter $\gamma$ is the fractional part of $1-\beta /4\pi$
such that $0\leq \gamma <1$.
We point out that for $\pi<\beta \leq 2\pi$
\begin{equation}
\label{10.5}
w_{4}(1-\frac{\beta}{4\pi})=-\frac{1}{720\pi^{2}}(-\frac{17}{8}+
\frac{45\pi}{\beta}-\frac{10\pi^{2}}{\beta^{2}}-\frac{16\pi^{4}}{\beta^{4}})
\end{equation}
and thus for $\beta =2\pi$, we have
\begin{equation}
\label{10.6}
w_{4}(\frac{1}{2})=-\frac{3}{128\pi^{2}}
\end{equation}
We have to take the limit of expression (\ref{10.4a}) when $\beta$ tends
to infinity. We can easily prove that
\begin{equation}
\label{10.7}
\lim_{\beta \rightarrow \infty}w_{4}(\gamma )=-\frac{11}{720\pi^{2}}
\end{equation}
For $\beta =2\pi$, by combining (\ref{10.6}) and (\ref{10.7}),
we recover (\ref{9.19}) from (\ref{10.4}).

\section{Conclusion}

Within the Euclidean approach to scalar and spinor field theory, we have
explicitly determined the Euclidean Green's functions in the Rindler
spacetime. Our results enable us to compute straightforwardly the vacuum
energy-momentum tensor relative to the Rindler observer, in particular
for massless fields in four dimensions.

However, in the latter case for a spinor field, we have found an
expression (\ref{9.19}) for the vacuum energy-momentum tensor which is
different to the one derived by two authors in previous papers
\cite{candelas2,dowker4,dowker1}.

\vspace*{1.5cm}{\large {\bf Acknowlegement}}

\vspace*{.5cm}I thank Dr L.A.J. London for reading the manuscript.

\end{document}